\documentclass[conference, 10pt,twoside,twocolumn]{IEEEtran}

\usepackage{cite}
\usepackage[T1]{fontenc}
\usepackage{graphicx}
\usepackage{amssymb}
\usepackage{amsmath}
\usepackage{amsthm}
\usepackage{subfigure}
\usepackage{booktabs} 
\usepackage{multirow}
\usepackage{microtype}
\usepackage{balance}
\usepackage{xcolor}
\usepackage{xfrac}    
\usepackage{algorithm}
\usepackage{setspace}
\usepackage{mdframed}
\usepackage{tikz}
\usepackage{circledsteps}
\usepackage{enumitem,kantlipsum}
\usepackage{twemojis}

\usepackage{algorithmicx}
\usepackage{algpseudocode}
\algrenewcommand\algorithmicindent{0.7em}%
\usepackage{xpatch}

\usepackage[hidelinks]{hyperref}

\usepackage{amssymb}
\usepackage{amsfonts}
\usepackage{mathrsfs}
\usepackage{xspace}
\usepackage{bm}
\usepackage{upgreek}

\newcommand{\safemath}[2]{\newcommand{#1}{\ensuremath{#2}\xspace}}

\safemath{\bma}{\mathbf{a}}
\safemath{\bmb}{\mathbf{b}}
\safemath{\bmc}{\mathbf{c}}
\safemath{\bmd}{\mathbf{d}}
\safemath{\bme}{\mathbf{e}}
\safemath{\bmf}{\mathbf{f}}
\safemath{\bmg}{\mathbf{g}}
\safemath{\bmh}{\mathbf{h}}
\safemath{\bmi}{\mathbf{i}}
\safemath{\bmj}{\mathbf{j}}
\safemath{\bmk}{\mathbf{k}}
\safemath{\bml}{\mathbf{l}}
\safemath{\bmm}{\mathbf{m}}
\safemath{\bmn}{\mathbf{n}}
\safemath{\bmo}{\mathbf{o}}
\safemath{\bmp}{\mathbf{p}}
\safemath{\bmq}{\mathbf{q}}
\safemath{\bmr}{\mathbf{r}}
\safemath{\bms}{\mathbf{s}}
\safemath{\bmt}{\mathbf{t}}
\safemath{\bmu}{\mathbf{u}}
\safemath{\bmv}{\mathbf{v}}
\safemath{\bmw}{\mathbf{w}}
\safemath{\bmx}{\mathbf{x}}
\safemath{\bmy}{\mathbf{y}}
\safemath{\bmz}{\mathbf{z}}
\safemath{\bmzero}{\mathbf{0}}
\safemath{\bmone}{\mathbf{1}}

\bmdefine{\biad}{a}
\bmdefine{\bibd}{b}
\bmdefine{\bicd}{c}
\bmdefine{\bidd}{d}
\bmdefine{\bied}{e}
\bmdefine{\bifd}{f}
\bmdefine{\bigd}{g}
\bmdefine{\bihd}{h}
\bmdefine{\biid}{i}
\bmdefine{\bijd}{j}
\bmdefine{\bikd}{k}
\bmdefine{\bild}{l}
\bmdefine{\bimd}{m}
\bmdefine{\bind}{n}
\bmdefine{\biod}{o}
\bmdefine{\bipd}{p}
\bmdefine{\biqd}{q}
\bmdefine{\bird}{r}
\bmdefine{\bisd}{s}
\bmdefine{\bitd}{t}
\bmdefine{\biud}{u}
\bmdefine{\bivd}{v}
\bmdefine{\biwd}{w}
\bmdefine{\bixd}{x}
\bmdefine{\biyd}{y}
\bmdefine{\bizd}{z}

\bmdefine{\bixid}{\xi}
\bmdefine{\bilambdad}{\lambda}
\bmdefine{\bimud}{\mu}
\bmdefine{\bithetad}{\theta}
\bmdefine{\biphid}{\phi}
\bmdefine{\bideltad}{\delta}

\safemath{\bmia}{\biad}
\safemath{\bmib}{\bibd}
\safemath{\bmic}{\bicd}
\safemath{\bmid}{\bidd}
\safemath{\bmie}{\bied}
\safemath{\bmif}{\bifd}
\safemath{\bmig}{\bigd}
\safemath{\bmih}{\bihd}
\safemath{\bmii}{\biid}
\safemath{\bmij}{\bijd}
\safemath{\bmik}{\bikd}
\safemath{\bmil}{\bild}
\safemath{\bmim}{\bimd}
\safemath{\bmin}{\bind}
\safemath{\bmio}{\biod}
\safemath{\bmip}{\bipd}
\safemath{\bmiq}{\biqd}
\safemath{\bmir}{\bird}
\safemath{\bmis}{\bisd}
\safemath{\bmit}{\bitd}
\safemath{\bmiu}{\biud}
\safemath{\bmiv}{\bivd}
\safemath{\bmiw}{\biwd}
\safemath{\bmix}{\bixd}
\safemath{\bmiy}{\biyd}
\safemath{\bmiz}{\bizd}

\safemath{\bmxi}{\bixid}
\safemath{\bmlambda}{\bilambdad}
\safemath{\bmmu}{\bimud}
\safemath{\bmtheta}{\bithetad}
\safemath{\bmphi}{\biphid}
\safemath{\bmdelta}{\bideltad}

\safemath{\bA}{\mathbf{A}}
\safemath{\bB}{\mathbf{B}}
\safemath{\bC}{\mathbf{C}}
\safemath{\bD}{\mathbf{D}}
\safemath{\bE}{\mathbf{E}}
\safemath{\bF}{\mathbf{F}}
\safemath{\bG}{\mathbf{G}}
\safemath{\bH}{\mathbf{H}}
\safemath{\bI}{\mathbf{I}}
\safemath{\bJ}{\mathbf{J}}
\safemath{\bK}{\mathbf{K}}
\safemath{\bL}{\mathbf{L}}
\safemath{\bM}{\mathbf{M}}
\safemath{\bN}{\mathbf{N}}
\safemath{\bO}{\mathbf{O}}
\safemath{\bP}{\mathbf{P}}
\safemath{\bQ}{\mathbf{Q}}
\safemath{\bR}{\mathbf{R}}
\safemath{\bS}{\mathbf{S}}
\safemath{\bT}{\mathbf{T}}
\safemath{\bU}{\mathbf{U}}
\safemath{\bV}{\mathbf{V}}
\safemath{\bW}{\mathbf{W}}
\safemath{\bX}{\mathbf{X}}
\safemath{\bY}{\mathbf{Y}}
\safemath{\bZ}{\mathbf{Z}}

\safemath{\bZero}{\mathbf{0}}
\safemath{\bOne}{\mathbf{1}}
\safemath{\bDelta}{\mathbf{\Delta}}
\safemath{\bLambda}{\mathbf{\UpLambda}}
\safemath{\bPhi}{\mathbf{\Upphi}}
\safemath{\bSigma}{\mathbf{\Upsigma}}
\safemath{\bOmega}{\mathbf{\Upomega}}
\safemath{\bTheta}{\mathbf{\Uptheta}}

\bmdefine{\biAd}{A}
\bmdefine{\biBd}{B}
\bmdefine{\biCd}{C}
\bmdefine{\biDd}{D}
\bmdefine{\biEd}{E}
\bmdefine{\biFd}{F}
\bmdefine{\biGd}{G}
\bmdefine{\biHd}{H}
\bmdefine{\biId}{I}
\bmdefine{\biJd}{J}
\bmdefine{\biKd}{K}
\bmdefine{\biLd}{L}
\bmdefine{\biMd}{M}
\bmdefine{\biOd}{N}
\bmdefine{\biPd}{O}
\bmdefine{\biQd}{P}
\bmdefine{\biRd}{R}
\bmdefine{\biSd}{S}
\bmdefine{\biTd}{T}
\bmdefine{\biUd}{U}
\bmdefine{\biVd}{V}
\bmdefine{\biWd}{W}
\bmdefine{\biXd}{X}
\bmdefine{\biYd}{Y}
\bmdefine{\biZd}{Z}

\bmdefine{\biDelta}{\Delta}
\bmdefine{\biLambda}{\Lambda}
\bmdefine{\biPhi}{\Phi}
\bmdefine{\biSigma}{\Sigma}
\bmdefine{\biOmega}{\Omega}
\bmdefine{\biTheta}{\Theta}

\safemath{\bimA}{\biAd}
\safemath{\bimB}{\biBd}
\safemath{\bimC}{\biCd}
\safemath{\bimD}{\biDd}
\safemath{\bimE}{\biEd}
\safemath{\bimF}{\biFd}
\safemath{\bimG}{\biGd}
\safemath{\bimH}{\biHd}
\safemath{\bimI}{\biId}
\safemath{\bimJ}{\biJd}
\safemath{\bimK}{\biKd}
\safemath{\bimL}{\biLd}
\safemath{\bimM}{\biMd}
\safemath{\bimN}{\biNd}
\safemath{\bimO}{\biOd}
\safemath{\bimP}{\biPd}
\safemath{\bimQ}{\biQd}
\safemath{\bimR}{\biRd}
\safemath{\bimS}{\biSd}
\safemath{\bimT}{\biTd}
\safemath{\bimU}{\biUd}
\safemath{\bimV}{\biVd}
\safemath{\bimW}{\biWd}
\safemath{\bimX}{\biXd}
\safemath{\bimY}{\biYd}
\safemath{\bimZ}{\biZd}

\safemath{\bimDelta}{\biDelta}
\safemath{\bimLambda}{\biLambda}
\safemath{\bimPhi}{\biPhi}
\safemath{\bimSigma}{\biSigma}
\safemath{\bimOmega}{\biOmega}
\safemath{\bimTheta}{\biTheta}

\safemath{\setA}{\mathcal{A}}
\safemath{\setB}{\mathcal{B}}
\safemath{\setC}{\mathcal{C}}
\safemath{\setD}{\mathcal{D}}
\safemath{\setE}{\mathcal{E}}
\safemath{\setF}{\mathcal{F}}
\safemath{\setG}{\mathcal{G}}
\safemath{\setH}{\mathcal{H}}
\safemath{\setI}{\mathcal{I}}
\safemath{\setJ}{\mathcal{J}}
\safemath{\setK}{\mathcal{K}}
\safemath{\setL}{\mathcal{L}}
\safemath{\setM}{\mathcal{M}}
\safemath{\setN}{\mathcal{N}}
\safemath{\setO}{\mathcal{O}}
\safemath{\setP}{\mathcal{P}}
\safemath{\setQ}{\mathcal{Q}}
\safemath{\setR}{\mathcal{R}}
\safemath{\setS}{\mathcal{S}}
\safemath{\setT}{\mathcal{T}}
\safemath{\setU}{\mathcal{U}}
\safemath{\setV}{\mathcal{V}}
\safemath{\setW}{\mathcal{W}}
\safemath{\setX}{\mathcal{X}}
\safemath{\setY}{\mathcal{Y}}
\safemath{\setZ}{\mathcal{Z}}
\safemath{\emptySet}{\varnothing}

\safemath{\colA}{\mathscr{A}}
\safemath{\colB}{\mathscr{B}}
\safemath{\colC}{\mathscr{C}}
\safemath{\colD}{\mathscr{D}}
\safemath{\colE}{\mathscr{E}}
\safemath{\colF}{\mathscr{F}}
\safemath{\colG}{\mathscr{G}}
\safemath{\colH}{\mathscr{H}}
\safemath{\colI}{\mathscr{I}}
\safemath{\colJ}{\mathscr{J}}
\safemath{\colK}{\mathscr{K}}
\safemath{\colL}{\mathscr{L}}
\safemath{\colM}{\mathscr{M}}
\safemath{\colN}{\mathscr{N}}
\safemath{\colO}{\mathscr{O}}
\safemath{\colP}{\mathscr{P}}
\safemath{\colQ}{\mathscr{Q}}
\safemath{\colR}{\mathscr{R}}
\safemath{\colS}{\mathscr{S}}
\safemath{\colT}{\mathscr{T}}
\safemath{\colU}{\mathscr{U}}
\safemath{\colV}{\mathscr{V}}
\safemath{\colW}{\mathscr{W}}
\safemath{\colX}{\mathscr{X}}
\safemath{\colY}{\mathscr{Y}}
\safemath{\colZ}{\mathscr{Z}}

\safemath{\opA}{\mathbb{A}}
\safemath{\opB}{\mathbb{B}}
\safemath{\opC}{\mathbb{C}}
\safemath{\opD}{\mathbb{D}}
\safemath{\opE}{\mathbb{E}}
\safemath{\opF}{\mathbb{F}}
\safemath{\opG}{\mathbb{G}}
\safemath{\opH}{\mathbb{H}}
\safemath{\opI}{\mathbb{I}}
\safemath{\opJ}{\mathbb{J}}
\safemath{\opK}{\mathbb{K}}
\safemath{\opL}{\mathbb{L}}
\safemath{\opM}{\mathbb{M}}
\safemath{\opN}{\mathbb{N}}
\safemath{\opO}{\mathbb{O}}
\safemath{\opP}{\mathbb{P}}
\safemath{\opQ}{\mathbb{Q}}
\safemath{\opR}{\mathbb{R}}
\safemath{\opS}{\mathbb{S}}
\safemath{\opT}{\mathbb{T}}
\safemath{\opU}{\mathbb{U}}
\safemath{\opV}{\mathbb{V}}
\safemath{\opW}{\mathbb{W}}
\safemath{\opX}{\mathbb{X}}
\safemath{\opY}{\mathbb{Y}}
\safemath{\opZ}{\mathbb{Z}}
\safemath{\opZero}{\mathbb{O}}
\safemath{\identityop}{\opI}

\safemath{\veca}{\bma}
\safemath{\vecb}{\bmb}
\safemath{\vecc}{\bmc}
\safemath{\vecd}{\bmd}
\safemath{\vece}{\bme}
\safemath{\vecf}{\bmf}
\safemath{\vecg}{\bmg}
\safemath{\vech}{\bmh}
\safemath{\veci}{\bmi}
\safemath{\vecj}{\bmj}
\safemath{\veck}{\bmk}
\safemath{\vecl}{\bml}
\safemath{\vecm}{\bmm}
\safemath{\vecn}{\bmn}
\safemath{\veco}{\bmo}
\safemath{\vecp}{\bmp}
\safemath{\vecq}{\bmq}
\safemath{\vecr}{\bmr}
\safemath{\vecs}{\bms}
\safemath{\vect}{\bmt}
\safemath{\vecu}{\bmu}
\safemath{\vecv}{\bmv}
\safemath{\vecw}{\bmw}
\safemath{\vecx}{\bmx}
\safemath{\vecy}{\bmy}
\safemath{\vecz}{\bmz}

\safemath{\veczero}{\bmzero}
\safemath{\vecone}{\bmone}
\safemath{\vecxi}{\bmxi}
\safemath{\veclambda}{\bmlambda}
\safemath{\vecmu}{\bmmu}
\safemath{\vectheta}{\bmtheta}
\safemath{\vecphi}{\bmphi}
\safemath{\vecdelta}{\bmdelta}

\safemath{\matA}{\bA}
\safemath{\matB}{\bB}
\safemath{\matC}{\bC}
\safemath{\matD}{\bD}
\safemath{\matE}{\bE}
\safemath{\matF}{\bF}
\safemath{\matG}{\bG}
\safemath{\matH}{\bH}
\safemath{\matI}{\bI}
\safemath{\matJ}{\bJ}
\safemath{\matK}{\bK}
\safemath{\matL}{\bL}
\safemath{\matM}{\bM}
\safemath{\matN}{\bN}
\safemath{\matO}{\bO}
\safemath{\matP}{\bP}
\safemath{\matQ}{\bQ}
\safemath{\matR}{\bR}
\safemath{\matS}{\bS}
\safemath{\matT}{\bT}
\safemath{\matU}{\bU}
\safemath{\matV}{\bV}
\safemath{\matW}{\bW}
\safemath{\matX}{\bX}
\safemath{\matY}{\bY}
\safemath{\matZ}{\bZ}
\safemath{\matzero}{\bmzero}

\safemath{\matDelta}{\bDelta}
\safemath{\matLambda}{\bLambda}
\safemath{\matPhi}{\bPhi}
\safemath{\matSigma}{\bSigma}
\safemath{\matOmega}{\bOmega}
\safemath{\matTheta}{\bTheta}

\safemath{\matidentity}{\matI}
\safemath{\matone}{\matO}

\safemath{\rnda}{A}
\safemath{\rndb}{B}
\safemath{\rndc}{C}
\safemath{\rndd}{D}
\safemath{\rnde}{E}
\safemath{\rndf}{F}
\safemath{\rndg}{G}
\safemath{\rndh}{H}
\safemath{\rndi}{I}
\safemath{\rndj}{J}
\safemath{\rndk}{K}
\safemath{\rndl}{L}
\safemath{\rndm}{M}
\safemath{\rndn}{N}
\safemath{\rndo}{O}
\safemath{\rndp}{P}
\safemath{\rndq}{Q}
\safemath{\rndr}{R}
\safemath{\rnds}{S}
\safemath{\rndt}{T}
\safemath{\rndu}{U}
\safemath{\rndv}{V}
\safemath{\rndw}{W}
\safemath{\rndx}{X}
\safemath{\rndy}{Y}
\safemath{\rndz}{Z}

\safemath{\rveca}{\bimA}
\safemath{\rvecb}{\bimB}
\safemath{\rvecc}{\bimC}
\safemath{\rvecd}{\bimD}
\safemath{\rvece}{\bimE}
\safemath{\rvecf}{\bimF}
\safemath{\rvecg}{\bimG}
\safemath{\rvech}{\bimH}
\safemath{\rveci}{\bimI}
\safemath{\rvecj}{\bimJ}
\safemath{\rveck}{\bimK}
\safemath{\rvecl}{\bimL}
\safemath{\rvecm}{\bimM}
\safemath{\rvecn}{\bimN}
\safemath{\rveco}{\bomO}
\safemath{\rvecp}{\bimP}
\safemath{\rvecq}{\bimQ}
\safemath{\rvecr}{\bimR}
\safemath{\rvecs}{\bimS}
\safemath{\rvect}{\bimT}
\safemath{\rvecu}{\bimU}
\safemath{\rvecv}{\bimV}
\safemath{\rvecw}{\bimW}
\safemath{\rvecx}{\bimX}
\safemath{\rvecy}{\bimY}
\safemath{\rvecz}{\bimZ}

\safemath{\rvecxi}{\bmxi}
\safemath{\rveclambda}{\bmlambda}
\safemath{\rvecmu}{\bmmu}
\safemath{\rvectheta}{\bmtheta}
\safemath{\rvecphi}{\bmphi}

\safemath{\rmatA}{\bimA}
\safemath{\rmatB}{\bimB}
\safemath{\rmatC}{\bimC}
\safemath{\rmatD}{\bimD}
\safemath{\rmatE}{\bimE}
\safemath{\rmatF}{\bimF}
\safemath{\rmatG}{\bimG}
\safemath{\rmatH}{\bimH}
\safemath{\rmatI}{\bimI}
\safemath{\rmatJ}{\bimJ}
\safemath{\rmatK}{\bimK}
\safemath{\rmatL}{\bimL}
\safemath{\rmatM}{\bimM}
\safemath{\rmatN}{\bimN}
\safemath{\rmatO}{\bimO}
\safemath{\rmatP}{\bimP}
\safemath{\rmatQ}{\bimQ}
\safemath{\rmatR}{\bimR}
\safemath{\rmatS}{\bimS}
\safemath{\rmatT}{\bimT}
\safemath{\rmatU}{\bimU}
\safemath{\rmatV}{\bimV}
\safemath{\rmatW}{\bimW}
\safemath{\rmatX}{\bimX}
\safemath{\rmatY}{\bimY}
\safemath{\rmatZ}{\bimZ}

\safemath{\rmatDelta}{\bimDelta}
\safemath{\rmatLambda}{\bimLambda}
\safemath{\rmatPhi}{\bimPhi}
\safemath{\rmatSigma}{\bimSigma}
\safemath{\rmatOmega}{\bimOmega}
\safemath{\rmatTheta}{\bimTheta}

\usepackage{amssymb}
\usepackage{amsfonts}
\usepackage{mathrsfs}
\usepackage{xspace}
\usepackage{bm}
\usepackage{fancyref}
\usepackage{textcomp}

\usepackage{multirow}
\usepackage{stmaryrd}

\newenvironment{textbmatrix}{	\setlength{\arraycolsep}{2.5pt}%
								\big[\begin{matrix}}{\end{matrix}\big]%
								\raisebox{0.08ex}{\vphantom{M}}}

\def\be{\begin{equation}}
\def\ee{\end{equation}}
\def\een{\nonumber \end{equation}}
\def\mat{\begin{bmatrix}}
\def\emat{\end{bmatrix}}
\def\btm{\begin{textbmatrix}}
\def\etm{\end{textbmatrix}}

\def\ba#1\ea{\begin{align}#1\end{align}}
\def\bas#1\eas{\begin{align*}#1\end{align*}}
\def\bs#1\es{\begin{split}#1\end{split}}
\def\bg#1\eg{\begin{gather}#1\end{gather}}
\def\bml#1\eml{\begin{multline}#1\end{multline}}
\def\bi#1\ei{\begin{itemize}#1\end{itemize}}

\newcommand{\lefto}{\mathopen{}\left}

\DeclareMathOperator{\Exop}{\opE}			

\newcommand{\orth}{\perp}					
\newcommand{\Ex}[2]{\ensuremath{\Exop_{#1}\lefto[#2\right]}} 	

\newcommand{\tp}[1]{\ensuremath{#1^{\text{T}}}} 		
\newcommand{\herm}[1]{\ensuremath{#1^{\text{H}}}} 	
\newcommand{\inv}[1]{\ensuremath{#1^{-1}}} 	
\newcommand{\pinv}[1]{\ensuremath{#1^{\dagger}}} 	

\safemath{\dirac}{\delta}					
\safemath{\krond}{\dirac}					

\safemath{\upto}{\uparrow}
\safemath{\downto}{\downarrow}
\safemath{\iu}{j}							
\safemath{\ev}{\lambda}						
\safemath{\hilseqspace}{l^{2}}				
\newcommand{\banachfunspace}[1]{\setL^{#1}}	
\safemath{\hilfunspace}{\banachfunspace{2}}	

\safemath{\SNR}{\textit{SNR}} 				
\safemath{\PAR}{\textit{PAR}} 				
\safemath{\No}{N_0}							
\safemath{\Es}{E_s}							
\safemath{\Eb}{E_b}							
\safemath{\EbNo}{\frac{\Eb}{\No}}
\safemath{\EsNo}{\frac{\Es}{\No}}

\DeclareMathOperator{\CHop}{\ensuremath{\opH}} 
\safemath{\tvir}{\rndh_{\CHop}}				
\safemath{\tvtf}{\rndl_{\CHop}}				
\safemath{\spf}{\rnds_{\CHop}}				
\safemath{\bff}{H_{\CHop}}					

\safemath{\ircf}{r_{h}}						
\safemath{\tftvcf}{r_{s}}					
\safemath{\tfcf}{r_{l}}						
\safemath{\bfcf}{r_{H}}						

\safemath{\tcorr}{c_h}						
\safemath{\scf}{c_{s}}						
\safemath{\tfcorr}{c_{l}}					
\safemath{\fcorr}{c_{H}}						

\safemath{\mi}{I}							
\safemath{\capacity}{C}						

\safemath{\normal}{\mathcal{N}}			
\safemath{\jpg}{\mathcal{CN}}			
\safemath{\mchain}{\leftrightarrow}		

\safemath{\dB}{\,\mathrm{dB}}
\safemath{\dBm}{\,\mathrm{dBm}}
\safemath{\Hz}{\,\mathrm{Hz}}
\safemath{\kHz}{\,\mathrm{kHz}}
\safemath{\MHz}{\,\mathrm{MHz}}
\safemath{\GHz}{\,\mathrm{GHz}}
\safemath{\s}{\,\mathrm{s}}
\safemath{\ms}{\,\mathrm{ms}}
\safemath{\mus}{\,\mathrm{\text{\textmu}s}}
\safemath{\ns}{\,\mathrm{ns}}
\safemath{\ps}{\,\mathrm{ps}}
\safemath{\meter}{\,\mathrm{m}}
\safemath{\mm}{\,\mathrm{mm}}
\safemath{\cm}{\,\mathrm{cm}}
\safemath{\m}{\,\mathrm{m}}
\safemath{\W}{\,\mathrm{W}}
\safemath{\mW}{\, \mathrm{mW}}
\safemath{\J}{\,\mathrm{J}}
\safemath{\K}{\,\mathrm{K}}
\safemath{\bit}{\,\mathrm{bit}}
\safemath{\nat}{\,\mathrm{nat}}

\safemath{\define}{\triangleq}			

\safemath{\equivalent}{\sim}
\safemath{\distas}{\sim}					
\safemath{\sdiff}{\Delta}				

\safemath{\reals}{\mathbb{R}}
\safemath{\positivereals}{\reals_{+}}
\safemath{\integers}{\mathbb{Z}}
\safemath{\posint}{\integers_{+}}
\safemath{\naturals}{\mathbb{N}}
\safemath{\posnaturals}{\naturals_{+}}
\safemath{\complexset}{\mathbb{C}}
\safemath{\rationals}{\mathbb{Q}}

\newcommand*{\fancyrefapplabelprefix}{app}		
\newcommand*{\fancyrefthmlabelprefix}{thm}		
\newcommand*{\fancyreflemlabelprefix}{lem}		
\newcommand*{\fancyrefcorlabelprefix}{cor}		
\newcommand*{\fancyrefdeflabelprefix}{def}		
\newcommand*{\fancyrefproplabelprefix}{prop}		
\newcommand*{\fancyrefexmpllabelprefix}{exmpl}
\newcommand*{\fancyrefalglabelprefix}{alg}		
\newcommand*{\fancyreftbllabelprefix}{tbl}		

\frefformat{vario}{\fancyrefseclabelprefix}{Section~#1}
\frefformat{vario}{\fancyrefthmlabelprefix}{Theorem~#1}
\frefformat{vario}{\fancyreftbllabelprefix}{Table~#1}
\frefformat{vario}{\fancyreflemlabelprefix}{Lemma~#1}
\frefformat{vario}{\fancyrefcorlabelprefix}{Corollary~#1}
\frefformat{vario}{\fancyrefdeflabelprefix}{Definition~#1}
\frefformat{vario}{\fancyreffiglabelprefix}{Fig.~#1}
\frefformat{vario}{\fancyrefapplabelprefix}{Appendix~#1}
\frefformat{vario}{\fancyrefeqlabelprefix}{(#1)}
\frefformat{vario}{\fancyrefproplabelprefix}{Proposition~#1}
\frefformat{vario}{\fancyrefexmpllabelprefix}{Example~#1}
\frefformat{vario}{\fancyrefalglabelprefix}{Algorithm~#1}

 \newtheorem{thm}{Theorem}
 
 \newtheorem{defi}{Definition}

 \newtheorem*{remark*}{Remark}

\safemath{\dictab}{[\,\dicta\,\,\dictb\,]}

\safemath{\ysig}{\bmy}
\safemath{\ysighat}{\hat{\ysig}}
\safemath{\ysigdim}{M}
\safemath{\xsig}{\bmx}
\safemath{\xsigdim}{N}
\safemath{\nx}{n_x}
\safemath{\zsig}{\bmz}
\safemath{\zsigdim}{\ysigdim}
\safemath{\rsig}{\bmr}
\safemath{\Adict}{\bA}
\safemath{\Adicttilde}{\widetilde{\Adict}}
\safemath{\Adictdim}{\outputdim\times\xsigdim}
\safemath{\avec}{\bma}
\safemath{\avectilde}{\tilde{\avec}}
\safemath{\Bdict}{\bB}
\safemath{\Bdicttilde}{\widetilde{\Bdict}}
\safemath{\Cdict}{\bC}
\safemath{\cvec}{\bmc}
\safemath{\Ddict}{\bD}
\safemath{\Ddictdim}{\ysigdim\times\xsigdim}
\safemath{\dvec}{\bmd}
\safemath{\Ddicttilde}{\widetilde{\bD}}
\safemath{\Bonb}{\bB}
\safemath{\bvec}{\bmb}
\safemath{\Bonbdim}{\ysigdim\times\ysigdim}
\safemath{\noise}{\bmn}
\safemath{\noisedim}{\ysigim}
\safemath{\err}{\bme}
\safemath{\errdim}{\ysigdim}
\safemath{\errset}{\setE}
\safemath{\nerr}{n_e}
\safemath{\delop}{\bP_\errset}
\safemath{\delopc}{\bP_{{\errset}^c}}

\safemath{\cplxi}{\imath}
\safemath{\cplxj}{\jmath}

\safemath{\dict}{\matD}
\safemath{\inputdim}{N}		
\safemath{\outputdim}{M}		
\safemath{\sparsity}{S}	
\safemath{\inputdimA}{{N_a}}	
\safemath{\inputdimB}{{N_b}}	
\safemath{\elemA}{{n_a}}	
\safemath{\elemB}{{n_b}}	
\safemath{\resA}{\matR_a}	
\safemath{\resB}{\matR_b}	
\safemath{\subD}{\matS} 
\safemath{\subA}{\matS_a} 
\safemath{\subB}{\matS_b} 
\safemath{\dicta}{\matA} 	
\safemath{\dictb}{\matB} 	
\safemath{\hollowS}{H}
\safemath{\hollowA}{H_a}
\safemath{\hollowB}{H_b}
\safemath{\cross}{Z}
\safemath{\coh}{\mu_d}			
\safemath{\coha}{\mu_a}			
\safemath{\cohb}{\mu_b}			
\safemath{\mubs}{\nu}	
\safemath{\cohm}{\mu_m} 
\safemath{\dictset}{\setD}	
\safemath{\dictsetp}{\dictset(\coh,\coha,\cohb)}	
\safemath{\dictsetgen}{\dictset_\text{gen}}
\safemath{\dictsetgenp}{\dictsetgen(\coh)}
\safemath{\dictsetonb}{\dictset_\text{onb}}
\safemath{\dictsetonbp}{\dictsetonb(\coh)}

\safemath{\leftside}{U}
\safemath{\rightsideA}{R_a}
\safemath{\rightsideB}{R_b}

\safemath{\indexS}{\setI_S} 

\safemath{\na}{n_a}			
\safemath{\nb}{n_b}			
\safemath{\coeffa}{p_i}	
\safemath{\coeffb}{q_j}	
\safemath{\seta}{\setP}		
\safemath{\setb}{\setQ}     
\safemath{\setw}{\setW}	
\safemath{\setz}{\setZ}	
\safemath{\cola}{\veca}		
\safemath{\colb}{\vecb}		
\safemath{\cold}{\vecd}		
\safemath{\inputvec}{\vecx} 	
\safemath{\error}{\vece}	
\safemath{\noiseout}{\vecz} 	
\safemath{\inputvecel}{x}
\safemath{\inputveca}{\vecx_a}
\safemath{\inputvecb}{\vecx_b}
\safemath{\outputvec}{\vecy}	
\safemath{\lambdamin}{\lambda_{\mathrm{min}}}

\safemath{\elltwo}{\ell_2}
\safemath{\ellone}{\ell_1}
\safemath{\ellzero}{\ell_0}
\safemath{\ellinf}{\ell_\infty}
\safemath{\ellinftilde}{\ell_{\widetilde\infty}}
\safemath{\licard}{Z(\coh,\coha,\cohb)}
\safemath{\xsol}{\hat{x}}
\safemath{\xbord}{x_b}		
\safemath{\xstat}{x_s}		
\safemath{\xstatLone}{\tilde{x}_s}
\safemath{\order}{\mathcal{O}} 
\safemath{\scales}{\Theta} 
\safemath{\ones}{\mathbf{1}} 
\safemath{\zeroes}{\mathbf{0}} 
\safemath{\thlone}{\kappa(\coh,\cohb)} 
\safemath{\constoneA}{\delta} 
\safemath{\constoneB}{\epsilon} 
\safemath{\nlarge}{L}				   
\safemath{\sumlarge}{S_\nlarge}
\safemath{\maxlarger}{P_\nlarge}	   
\safemath{\Pzero}{\textrm{P0}}	
\safemath{\Pone}{\textrm{P1}}
\safemath{\vecfir}{\vecw}			 
\safemath{\vecsec}{\vecz}
\safemath{\elvecfir}{w}              
\safemath{\elvecsec}{z}				 
\safemath{\nlargefir}{n}
\safemath{\normout}{\gamma}
\safemath{\auxfun}{h}
\safemath{\supp}{\textrm{supp}}

\safemath{\indexa}{\ell}
\safemath{\indexb}{r}
\safemath{\indexc}{i}
\safemath{\indexd}{j}

\safemath{\project}{P}

\usepackage{framed}

\renewcommand{\bSigma}{\boldsymbol{\Sigma}}
\newcommand{\bsigma}{\boldsymbol{\sigma}}
\newcommand{\Ie}{I^{\boldsymbol{\ast}}}

\newcommand{\secret}{\makebox[7pt][l]{\raisebox{-0.03cm}[0pt][0pt]{\twemoji[height=2.5mm,trim={0.1mm 0.1mm 0.1mm 0mm}, clip]{game_die}}}}
\newcommand{\largesecret}{\twemoji[height=3mm,trim={0.1mm 0.1mm 0.1mm 0mm}, clip]{game_die}}

\safemath{\bsfU}{\boldsymbol{\mathsf{U}}}
\safemath{\bsfV}{\boldsymbol{\mathsf{V}}}
\safemath{\bsfSigma}{\boldsymbol{\mathsf{\Sigma}}}
\safemath{\Cpar}{\bC_{\parallel}}
\safemath{\Corth}{\bC_{\orth}}
\safemath{\Ypar}{\bY_{\parallel}}
\safemath{\Yorth}{\bY_{\orth}}

\IEEEoverridecommandlockouts
\allowdisplaybreaks 

\newcommand*\tinygraycircled[1]{\Circled[inner color=white, fill color= gray, outer color=gray]{\footnotesize{\textnormal{#1}}}}
\newcommand*\scriptsizegraycircled[1]{\Circled[inner color=white, fill color= gray, outer color=gray]{\scriptsize{\textnormal{#1}}}}

\safemath{\Hj}{\bJ}
\safemath{\bsj}{\bmw}
\safemath{\sj}{w}
\safemath{\Ej}{E_w}
\safemath{\proxg}{\text{prox}_g}
\safemath{\rE}{\rho_{\textsf{\tiny{E}}}}
\safemath{\rP}{\rho_{\textsf{\tiny{P}}}}

\begin{document}
\bstctlcite{IEEEexample:BSTcontrol} 

\title{Universal MIMO Jammer Mitigation via\\ Secret Temporal Subspace Embeddings}

\author{\IEEEauthorblockN{Gian Marti and Christoph Studer}\\
\IEEEauthorblockA{\em Department of Information Technology
and Electrical Engineering, ETH Zurich, Switzerland\\
email: gimarti@ethz.ch and studer@ethz.ch
}
\thanks{The work of CS was supported in part by the U.S. National Science Foundation (NSF) 
under grants CNS-1717559 and ECCS-1824379. The work of GM and CS was supported in part by an ETH Research Grant.}
\thanks{Emojis by Twitter, Inc. and other contributors are licensed under CC-BY~4.0.}
}

\maketitle

\begin{abstract}

MIMO processing enables jammer mitigation through spatial filtering, provided 
that the receiver knows the spatial signature of the jammer interference.
Estimating this signature is easy for barrage jammers that transmit continuously and with static 
signature, 
but difficult for more sophisticated jammers: Smart jammers may deliberately suspend transmission
when the receiver tries to estimate their spatial signature, they may use time-varying beamforming
to continuously change their spatial signature, or they may stay mostly silent and jam only  
specific instants (e.g., transmission of control signals).
To deal~with such smart jammers, we propose MASH, the first method that indiscriminately mitigates \emph{all} types of jammers: Assume that the transmitter and receiver share a common secret. 
Based on this secret, the transmitter \emph{embeds} (with a linear time-domain transform) 
its signal in a secret subspace of a higher-dimensional space.
The receiver applies a reciprocal linear transform to the receive signal, which (i) \emph{raises} the legitimate transmit signal from its secret subspace 
and (ii) provably transforms \emph{any} jammer into a barrage 
jammer, which makes estimation and mitigation via MIMO processing straightforward. 
We show the efficacy of MASH for data transmission in the massive multi-user MIMO~uplink.

\end{abstract}

\section{Introduction}
Jammers are a critical threat to wireless communication systems, which have become indispensable to  society\mbox{\cite{threatvectors2021cisa, idr2022unmanned, economist2021satellite}.}
Techniques such as direct-sequence spread spectrum~\cite{guanella1944dsss, madhow1994mmse}
or frequency-hopping spread spectrum\mbox{\cite{tesla1903fhss, stark1985coding}} provide a certain degree of robustness against interference, but are no match for strong wideband jammers. 
MIMO processing, in contrast, is able to completely remove jammer interference 
through spatial filtering~\cite{leost2012interference} and is therefore a promising path towards jammer-resilient communications.
However, MIMO jammer mitigation relies on accurate knowledge of the jammer's spatial 
signature (e.g., its subspace or its spatial covariance matrix) at the receiver. 
This signature is easy to estimate for barrage jammers (jammers that transmit continuously 
and with time-invariant signature):
The receiver can analyze the receive signal from a training period
which provides a representative snapshot of the jammer's signature. 
A more sophisticated jammer, however, may thwart such a na\"ive approach, for example by deliberately suspending jamming
during the training period, by manipulating its spatial signature through time-varying beamforming, 
or by jamming only specific communication parts (e.g., control signals) and staying unnoticeable during other parts
\cite{ miller2010subverting,lichtman2016communications,lichtman20185g}. 
In all such cases, the receive signal during a training period does \emph{not} 
pro-vide a representative snapshot of the jammer's spatial signature.

\subsection{Contributions}
We propose MASH (short for MitigAtion via Subspace Hiding), 
a novel approach for MIMO jammer mitigation, and the first that mitigates \emph{all} types of jammers, no 
matter how ``smart'' they are. 
MASH assumes that the transmitter(s) and the receiver share a common secret. 
Based on this secret, the transmitter(s) embed their time-domain signals in a secret subspace of a higher-dimensional 
space using a linear time-domain transform, and the receiver applies a reciprocal linear time-domain transform to its receive signal. We show that the receiver's transform
(i)~\emph{raises} the legitimate transmit signal from its secret subspace 
and (ii)~provably transforms \emph{any} jammer into a barrage jammer, which renders it easy to mitigate.
To showcase the effectiveness of MASH, we consider data transmission in the massive multi-user MIMO uplink. 
We provide three exemplary variations of MASH for this scenario, and we 
highlight its universality through extensive simulations. 

\subsection{State of the Art}

The potential of MIMO processing for jammer mitigation is well known \cite{pirayesh2022jamming}. 
A variety of methods have been proposed that aim at mitigating barrage jammers 
\cite{marti2021hybrid, jiang2021efficient, chehimi2023machine, marti2021snips, yang2022estimation, he2022high, do18a, akhlaghpasand20a, akhlaghpasand20b, zeng2017enabling}, which either assume perfect knowledge of the jammer's spatial signature at the receiver 
\cite{marti2021hybrid, jiang2021efficient, chehimi2023machine} 
or exploit the jammer's stationarity by using a  training period 
\cite{marti2021snips, yang2022estimation, he2022high, do18a, akhlaghpasand20a, akhlaghpasand20b, zeng2017enabling}:
References\mbox{\cite{marti2021snips, yang2022estimation, he2022high}} estimate the jammer's spatial signature during a  dedicated
training period in which the legitimate transmitters are idle. 
References \cite{do18a, akhlaghpasand20a} estimate the jammer's spatial signature during a prolonged
pilot phase in which the jammer interference is separated from the legitimate transmit signals by projecting the receive
signals onto an unused pilot sequence that is orthogonal to the legitimate pilots. 
References~\cite{zeng2017enabling, akhlaghpasand20b} propose a jammer-resilient data detector directly as a function of certain spatial filters which are estimated during the pilot~phase. 
While these methods are only effective against barrage jammers, MASH first transforms \emph{any}
jammer into a barrage jammer and therefore can be used to improve the efficacy of the aforementioned methods 
for the mitigation of non-barrage~jammers. 

It is widely known that jammers do not need to transmit in a time-invariant manner.
Several MIMO methods have therefore been proposed for mitigating time-varying jammers
\cite{shen14a, yan2016jamming, hoang2021suppression, hoang2022multiple, marti2023maed, marti2023jmd}. 
References \cite{shen14a, yan2016jamming} consider a reactive jammer that only jams when the legitimate
transmitters are actively transmitting. Such methods mitigate the jammer by using training periods in which the legitimate
transmitters send pre-defined symbols that can be compensated for at the receiver. 
References\mbox{\cite{hoang2021suppression, hoang2022multiple}} assume a multi-antenna jammer that continuously
varies its subspace through dynamic beamforming. Such methods mitigate the jammer using training periods (during which the
legitimate transmitters are idle) recurrently and whenever the jammer subspace is believed to 
have changed substantially. 
References \cite{marti2023maed, marti2023jmd} consider general smart jammers. To mitigate such jammers, they  
propose a novel paradigm called joint jammer mitigation and data detection (JMD) in which the jammer subspace
is estimated and nulled jointly with detecting the transmit data over an entire communication frame. 

All of these methods make stringent assumptions about the jammer's transmit behavior
(i.e., that the jammer is a barrage \cite{zeng2017enabling, jiang2021efficient, yang2022estimation, he2022high, chehimi2023machine, marti2021hybrid, marti2021snips, do18a, akhlaghpasand20a, akhlaghpasand20b} 
or reactive jammer \cite{shen14a, yan2016jamming}, or that its beamforming changes only slowly \cite{hoang2021suppression, hoang2022multiple}), 
or they require solving a hard optimization problem while still being susceptible to certain types of jammers \cite{marti2023maed, marti2023jmd}.
Moreover, the number of jammer antennas is often required to be known in advance (see e.g., \cite{yan2016jamming, akhlaghpasand20a , marti2023jmd}).
MASH, in contrast, makes \emph{no} assumptions about the jammer's transmit behavior, 
does \emph{not} require solving an optimization problem, is effective against \emph{all} 
jammer types, and does \emph{not} require the number of jammer antennas to be known in advance.

\subsection{Notation}
Matrices and column vectors are represented by boldface uppercase and lowercase letters, respectively.
For a matrix~$\bA$, the transpose is $\tp{\bA}$, the conjugate transpose is $\herm{\bA}$, 
the submatrix consisting of the columns (rows) from $n$ through~$m$ is $\bA_{[n:m]}$ ($\bA_{(n:m)}$),
and the Frobenius norm is $\| \bA \|_F$.
The columnspace and rowspace of $\bA$ are $\text{col}(\bA)$ and $\text{row}(\bA)$, respectively.
Horizontal concatenation of two matrices $\bA$ and~$\bB$ is denoted by $[\bA,\bB]$; vertical concatenation is $[\bA;\bB]$.
The $k$th column and row of $\bA$ are $\bma_k$ and $\tp{\bma_{(k)}}$, respectively. 
The $N\!\times\!N$ identity matrix is $\bI_N$.
The compact singular value decomposition (SVD) of a rank-$R$ matrix $\bA\in\opC^{M\times N}$ is denoted
$\bA=\bsfU\bsfSigma\herm{\bsfV}$, where $\bsfU\in\opC^{M\times R}$ and $\bsfV\in\opC^{N\times R}$~have orthonormal 
columns and $\bsfSigma\in\opR^{R\times R}$ is a diagonal matrix whose entries on the main diagonal are the non-zero singular 
values of $\bA$ in decreasing order. The main diagonal of $\bsfSigma$ is denoted by $\bsigma=\text{diag}(\bsfSigma)$.

\section{System Model}\label{sec:setting}

We consider jammer mitigation in communication systems with multi-antenna receivers. 
For ease of exposition, we focus on the multi-user (MU) MIMO uplink, but we emphasize
that our method can also be applied in single-user (SIMO) or point-to-point MIMO settings.
We consider a frequency-flat transmission model\footnote{An extension to frequency-selective channels 
with OFDM is possible \cite{marti2023single}.}
with the following input-output relation:
\begin{align}
	\bmy_k = \bH \bmx_k + \Hj \bsj_k + \bmn_k.  \label{eq:model}
\end{align}
Here, $\bmy_k\in\opC^B$ is the time-$k$ receive vector at the basestation (BS),
$\bH \in \opC^{B\times U}$ is the channel matrix of $U$ legitimate single-antenna user equipments (UEs) 
with time-$k$ transmit vector $\bmx_k\in\opC^U$,
$\bJ \in\opC^{B\times I}$ is the channel matrix of an \mbox{$I$-antenna} jammer
(or of multiple jammers, with $I$ being the total number of jammer antennas)
with time-$k$ transmit vector \mbox{$\bmw_k\in\opC^{I}$},
and $\bmn_k\sim\setC\setN(\mathbf{0},\No\bI_B)$ is i.i.d.\ circularly-symmetric complex white Gaussian  noise with 
per-entry variance~$\No$.

We assume that the jammer can vary its transmit characteristics.
For this, we use a model in which the jammer transmits 
\begin{align}
	\bsj_k = \bA_k \tilde\bsj_k
\end{align}
at time $k$, where, without loss of generality, 
$\Ex{}{\tilde\bmw_k\herm{\tilde\bmw_k}}=\bI_I$ for all $k$,
and where $\bA_k\in\opC^{I\times I}$ is a beamforming matrix that can change \emph{arbitrarily} as a function of~$k$. 
In particular,~$\bA_k$ can be the all-zero matrix (the jammer suspends jamming at time $k$), 
some of its rows can be zero (the jammer uses only a subset of its antennas at time~$k$), 
or it can be rank-deficient in another~way. 
Moreover, there need not be any relation between~$\bA_k$ and $\bA_{k'}$ for~$k\neq k'$.
Our only assumption is that the number of jammer antennas $I$ is smaller than the number of BS antennas $B$.

Our proposed method operates on transmission frames of length $L$, and we assume that the
channels $\bH$ and $\bJ$ stay approximately constant for the duration of such a frame. 
We therefore restate~\eqref{eq:model} for a length-$L$ transmission frame as
\begin{align}
	\bY = \bH\bX + \bJ\bW + \bN, \label{eq:blockmodel}
\end{align}
where $\bY=[\bmy_1,\dots,\bmy_L]\in\opC^{B\times L}$ is the receive matrix, and
$\bX=[\bmx_1,\dots,\bmx_L]\in\opC^{U\times L}$~and $\bW = [\bA_1\tilde\bmw_1,\dots,\bA_L\tilde\bmw_L]\in\opC^{I\times L}$ are the corresponding 
transmit matrices. Note that, since there are no restrictions for the matrices $\bA_k$, 
$\bW$ can be arbitrary and need in general not follow an explicit probabilistic model.

\section{On Barrage Jammers} \label{sec:barrage}
We prepare the ground for the exposition of MASH by considering a
class of jammers that is easy to mitigate: \emph{barrage jammers}, 
which transmit noise-like stationary signals over the entire spectrum \cite{lichtman2016communications}. 
In general, the notion of barrage jammers neither includes nor excludes the use of transmit beamforming. 

We now provide a novel, formal notion of barrage jammers. This notion explicitly 
allows for transmit beamforming,
as long as the beams stay constant for the duration of a communication~frame. 
Following our model in \fref{sec:setting}, we consider communication in frames of length $L$. 
During such a frame, a jammer transmits some matrix $\bW\in\opC^{I\times L}$ 
with corresponding receive interference $\bJ\bW \in \opC^{B\times L}$. 
We denote by $\Ie$ the rank of~$\bJ\bW$, i.e., $\Ie$ is the dimension of the interference space
which, depending on the jammer's transmit beamforming type, can be equal to or strictly smaller than $I$. 
We now consider the compact SVD of the interference, which we write as
\begin{align}
 	\bJ\bW = \bsfU\bsfSigma\herm{\bsfV}. \label{eq:barrage_svd}
\end{align}
We call $\bsfU\in\opC^{B\times \Ie}$ the \emph{spatial scope} of the jammer: 
The columns of this matrix form an orthonormal basis of the interference space in the spatial domain.   
We call ${\bsfV}\in\opC^{L\times\Ie}$ the \emph{temporal extension} of the jammer within the frame: 
its columns form an orthonormal basis of the interference space in the time domain. 
Finally, we call $\bsigma = \text{diag}(\bsfSigma)=\tp{[\sigma_1,\dots,\sigma_{\Ie}]}$ (with $\sigma_1\geq\dots\geq\sigma_{\Ie}>0$) the \emph{energy profile}, which 
determines how much energy the jammer allocates to the different dimensions in space and time. 
While the decomposition of the receive interference in \eqref{eq:barrage_svd} does not assume a probabilistic jammer, 
our definition of barrage jammers requires probabilistic behavior: \vspace{-1mm}

\begin{defi} \label{def:barrage}
A \emph{barrage jammer} is a jammer for which the columns of $\bsfV$ are distributed uniformly over the complex \mbox{$L$-dimensional} unit sphere.
\end{defi}
Barrage jammers are fully characterized by their spatial scope $\bsfU$ and their energy profile $\bsigma$.
Examples are a jammer that transmits $\bW$ with i.i.d. $\setC\setN(0,1)$ entries
or a jammer that transmits $\bW=\bma\tp{\bmw}$, with 
$\bma\in\opC^I$ and with $\bmw\sim\setC\setN(\mathbf{0},\bI_B)$. 

\begin{figure}[tp]
\centering
\includegraphics[height=3.75cm]{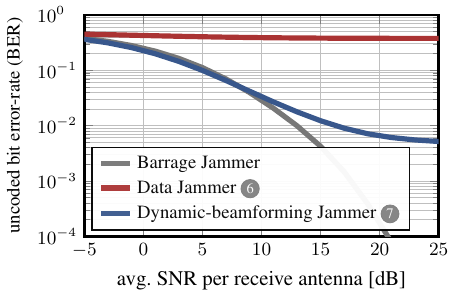}
\caption{
Error rate performance of a coherent communication receiver that~uses a training  period-based orthogonal projection 
for mitigating the jammer, least squares channel estimation, and LMMSE data detection.
The setup is described in \fref{sec:eval} and the $10$-antenna jammers are a barrage jammer transmitting 
i.i.d. $\setC\setN(0,1)$ symbols as well as the jammers $\scriptsizegraycircled{6}$ and $\scriptsizegraycircled{7}$
from \fref{sec:jammers}.
}
\label{fig:different_jammers}
\vspace{-2mm}
\end{figure}

Such a formal definition of barrage jammers allows us to see why 
barrage jammers can be mitigated easily and effectively: 
Consider a length-$L$ communication frame as in~\eqref{eq:blockmodel}, where the jammer is a barrage
jammer with compact SVD as given in~\eqref{eq:barrage_svd}. For simplicity, let 
us neglect the impact of thermal noise by assuming $\bN=\mathbf{0}$. 
If we insert a jammer training period of duration $R$ at the start of the frame, 
i.e., if the first $R$ columns of $\bX$ are zero, then the corresponding receive signal is 
\begin{align}
	\bY_{[1:R]} = \bJ\bW_{[1:R]} = \bsfU\bsfSigma(\herm{\bsfV})_{[1:R]}. \label{eq:train}
\end{align}
Since the columns of $\bsfV$ are uniformly distributed over the complex $L$-dimensional unit sphere, 
the truncated temporal extension $(\herm{\bsfV})_{[1:R]}\in\opC^{\Ie\times R}$ 
has full rank $\min\{\Ie,R\}$ with probability one. 
So provided that $\Ie\geq R$, we have $\text{col}(\bY_{[1:R]})=\text{col}(\bsfU)$ with probability one. 
The receiver can thus find the jammer subspace $\text{col}(\bJ\bW)$ \emph{of the entire frame}
from the compact SVD $\bY_{[1:R]}\!=\!\tilde\bsfU\tilde\bsfSigma\herm{\tilde\bsfV}$ as 
$\text{col}(\bJ\bW)\!=\!\text{col}(\tilde\bsfU)$,
and it can mitigate the jammer \emph{for the entire frame} using a projection matrix 
$\bP=\bI_B - \tilde\bsfU\herm{\tilde\bsfU}$, which satisfies $\bP\bJ\bW=\mathbf{0}_{B\times L}$. 

A smart or dynamic jammer, in contrast, might stop jamming during these $R$ training samples, 
in which case \eqref{eq:train} becomes \mbox{$\bY_{[1:R]} =\mathbf{0}_{B\times R}$} so that
the projection \mbox{$\bP=\bI_B$} is simply the identity, which will not mitigate
any subsequent interference of the jammer, since $\bP\bJ\bW_{[R+1:L]}=\bJ\bW_{[R+1:L]}$. 

Similarly, a dynamic multi-antenna jammer could at any given instant use only a subset 
of its antennas but~switch between different subsets over time, thereby changing the~spatial 
subspace of its interference. If $\text{col}(\bW_{[1:R]})\not\supseteq\text{col}(\bW_{[R+1:L]})$
and the columns of $\bJ$ are linearly independent, then 
$\text{col}(\bJ\bW_{[1:R]})\not\supseteq\text{col}(\bJ\bW_{[R+1:L]})$,
so the projection $\bP$ will not completely mitigate the jammer, 
$\bP\bJ\bW_{[R+1:L]}\neq\mathbf{0}_{B\times(L-R)}$.

\fref{fig:different_jammers} shows the effectiveness of jammer mitigation with~an orthogonal projection based on a 
training period for a barrage jammer as well as for two non-barrage jammers 
(the simulations of \fref{fig:different_jammers} \emph{do} take into account the effects of thermal 
noise $\bN$).

\section{MASH: Universal MIMO Jammer Mitigation\\ via Secret Temporal Subspace Embeddings} \label{sec:mash}
We are now ready to present MASH.
We assume that the UEs and the BS share a common secret~\secret.
Based on this secret, the UEs and the BS construct, in pseudo-random manner, 
a unitary matrix $\bC=f(\secret)$ which is uniformly (or \emph{Haar}) distributed over the set of all unitary $L\times L$
matrices.\footnote{See \cite[Sec.~1.2]{meckes2019random} on how to construct Haar distributed matrices.}
Since the jammer does not know \secret, it does not know $\bC$.
The matrix $\bC$ is then divided row-wise into two submatrices $\bC=[\Corth; \Cpar]$ 
with $\Corth\in\opC^{R\times L}$ and $\Cpar^{K \times L}$, 
where~$R$ and~$K$ are non-negative integers such that $R+K=L$. We call $R$ the \emph{redundancy}, and we require
that $R\geq \Ie$, i.e., that the redundancy is at least as big as the rank of the interference.\footnote{We do not require this dimension to be known a priori. The choice 
 of $R$ simply limits number of interference dimensions that can be mitigated.}

The UEs use the matrix $\Cpar$ to \emph{embed} a \mbox{length-$K$} transmit signal $\bS\in\opC^{U\times K}$ 
in the secret $K$-dimensional subspace of~$\opC^L$ that is spanned by the rows of $\Cpar$ by transmitting
\begin{align}
	\bX = \bS\Cpar. \label{eq:embed}
\end{align}
The transmit signal $\bS$ could consist, e.g., of pilots for channel estimation, control signals, or data symbols. 
Embedding $\bS$ into the secret subspace requires no cooperation between the UEs: The $u$th UE simply transmits the 
$u$th row of $\bX$, which it can compute as $\tp{\bmx_{(u)}}=\tp{\bms_{(u)}}\Cpar$, 
where $\tp{\bms_{(u)}}$ is its own transmit signal. 
With $\bX$ as in \eqref{eq:embed}, the receive matrix in \eqref{eq:blockmodel}~becomes 
\begin{align}
	\bY &= \bH \bX + \bJ\bW + \bN \label{eq:physical_io} \\
	&=\bH \bS \Cpar + \bJ\bW + \bN. \label{eq:mash_io}
\end{align}
Since $\bC$ is unknown to the jammer, $\bW$ is independent of~$\bC$. 
(This does not imply a probabilistic model for $\bW$---it only means that $\bW$ is no function of $\bC$.)
We make no other assump-tions about $\bW$. 
In particular, we also allow that $\bW$ might depend on $\bH$ and $\bJ$ (the jammer has full channel knowledge), 
or even on~$\bS$ (the jammer knows the signal to be transmitted).

Having received $\bY$ as in \eqref{eq:mash_io}, 
the receiver \emph{raises} the signals by multiplying $\bY$ from the right
with $\herm\bC$, obtaining
\begin{align}
	\bar{\bY} &\triangleq \bY\herm{\bC}
	= \bH \bS \Cpar\herm{\bC} 
	 + \bJ\underbrace{\bW\herm{\bC}}_{~~\triangleq \bar{\bW}\!\!}
	 + \underbrace{\bN\herm{\bC}}_{~\,\triangleq \bar{\bN}\!\!} \\
	&= [\mathbf{0}_{B\times R}, \bH\bS] + \bJ\bar{\bW} + \bar{\bN}, \label{eq:raised}
\end{align}
where $\Cpar\herm{\bC}=[\mathbf{0}_{K\times R}, \bI_K]$ in \eqref{eq:raised} follows from the 
unitarity of $\bC=[\Corth; \Cpar]$. 
We obtain an input-output relation in which the channels $\bH$ and $\bJ$
are unchanged, the UEs are~idle for the first $R$ samples and transmit~$\bS$ in the remaining $K$ samples, 
the jammer transmits $\bar{\bW} = \bW\herm{\bC}$, and the noise is \mbox{$\bar{\bN} = \bN\herm{\bC}$} and
is i.i.d. circularly-symmetric complex Gaussian with variance~$\No$.
We call the domain of the input-output relation \eqref{eq:raised} the \emph{message domain}, 
which contrasts with the \emph{physical domain} of \eqref{eq:mash_io}. 
We now state our key theoretical result, the proof of which is in \fref{app:app}: 
\begin{thm} \label{thm:barrage}
	Consider the decomposition of the physical domain jammer~interference $\bJ\bW$ in \eqref{eq:mash_io}
	into its spatial scope~$\bsfU$, its temporal extension $\bsfV$, and  its energy profile $\bsigma$. 
	Then the jammer interference $\bJ\bar{\bW}$ in the message domain \eqref{eq:raised} has identical spatial scope
	$\bar{\bsfU}=\bsfU$ and energy profile $\bar{\bsigma}=\bsigma$, 	but~its temporal extension $\bar{\bsfV}$ is a random 
	matrix whose columns are uniformly distributed over the complex $L$-dimensional unit sphere.
	Thus, $\bJ\bar{\bW}$ is the receive interference of a barrage jammer. 
\end{thm}

In other words, MASH transforms \emph{any} jammer into the (unique) barrage jammer 
with identical spatial scope and identical energy profile.
Furthermore, we can see in the message domain input-output relation \eqref{eq:raised} that 
MASH also conveniently deals with the legitimate transmit signals: 
The first~$R$ columns of the message domain receive signal $\bar{\bY}$ 
contain no UE signals and thus correspond 
to a jammer training period which can be used for 
constructing a jammer-mitigating filter that is effective for the entire frame (since the
message domain jammer is a barrage jammer).
The remaining $K$ columns of $\bar{\bY}$ correspond to the transmit signal $\bS$, which can be recovered 
using the jammer-mitigating filter from the training period. 

To concretize MASH, we consider the case of coherent data transmission,
for which we propose three different ways to proceed from~\fref{eq:raised}: 
(i)~orthogonal projection, (ii)~LMMSE equalization, and (iii)~joint jammer mitigation and data detection. 
In coherent data transmission, the transmit signal $\bS=[\bS_T,\bS_D]$ consists of 
orthogonal pilots $\bS_T\in\opC^{U\times U}$ and data symbols $\bS_D\in\setS^{U\times (K-U)}$ taken from 
a constellation $\setS$.
To simplify the ensuing discussion, we define $[\bar{\bY}_J,\bar{\bY}_T,\bar{\bY}_D]\triangleq\bar{\bY}$ 
and rewrite \eqref{eq:raised} as three separate equations:
\vspace{-1mm}
\begin{align}
	\bar{\bY}_J &= \bJ\bar{\bW}_{J} + \bar{\bN}_J \in\opC^{B\times R} \label{eq:j_train}\\
	\bar{\bY}_T &= \bH\bS_T + \bJ\bar{\bW}_{T} + \bar{\bN}_T \in\opC^{B\times U} \label{eq:pilots}\\
	\bar{\bY}_D &= \bH\bS_D + \bJ\bar{\bW}_{D} + \bar{\bN}_D	 \in\opC^{B\times (K-U)}. \label{eq:data}
\end{align}
\vspace{-8mm}
\subsection{Orthogonal Projection} \label{sec:mash-pos}
The matrix $\bar{\bY}_J$ contains samples of a barrage jammer corrupted by white Gaussian noise $\bar{\bN}_J$, 
but not by any UE signals. We can thus use $\bar{\bY}_J$ to estimate the projection onto the orthogonal complement of the jammer's
spatial scope (cf.~\fref{sec:barrage}).
The maximum-likelihood estimate of the jammer's spatial scope is given by the $\Ie$ leading left-singular vectors
$\bU=[\bmu_1,\dots,\bmu_{\Ie}]\in\opC^{U\times \Ie}$ of $\bar{\bY}_J$.\footnote{The effective dimension $\Ie$ of
the jammer interference can directly be estimated from $\bar{\bY}_J$ as the number of its singular values
that significantly exceed the threshold $\sqrt{B\No}$ which is expected due to thermal noise.
\label{footnote:est_Ie}}
The corresponding projection matrix is therefore 
\begin{align}
	\hat\bP=\bI_B - \bU\herm{\bU}, \label{eq:mash_barrage_proj}
\end{align} 
and the jammer can be mitigated in the pilot and data phase~via
\begin{align}
	\bar\bY_{\bP,T} &\triangleq \hat\bP\bar{\bY}_T = \underbrace{\hat\bP\bH}_{~\triangleq\bH_\bP\!\!\!\!\!\!\!}\bS_T 
	+ \hat\bP\bJ\bar{\bW}_{T} + \underbrace{\hat\bP\bar{\bN}_{T}}_{~~~\,\triangleq \bN_{\bP,T}\!\!\!\!\!\!} \\
	&\approx \bH_\bP\bS_T + \bN_{\bP,T}, \label{eq:pos_chest}
\end{align} 
and
\begin{align}
	\bar\bY_{\bP,D} &\triangleq \hat\bP\bar{\bY}_D = \hat\bP\bH\bS_D + \hat\bP\bJ\bar{\bW}_{D} + \hat\bP\bar{\bN}_{D} \\
	&\approx \bH_\bP\bS_D + \bN_{\bP,D}.	 \label{eq:pos_det}
\end{align}
where the approximations in \eqref{eq:pos_chest} and \eqref{eq:pos_det} hold because
$\hat\bP\bJ\bar{\bW}_{T}\approx\mathbf{0}$ and $\hat\bP\bJ\bar{\bW}_{D}\approx\mathbf{0}$.
The BS thus obtains an input-output relation that is jammer-free and consists of a virtual channel $\bH_\bP$
corrupted by~Gaussian noise with spatial distribution $\setC\setN(\mathbf{0},\No\bP)$.
The receiver can now estimate the virtual channel $\bH_\bP$, e.g., using least-squares (LS),~as follows:
\begin{align}
	\hat\bH_\bP = \bar\bY_{\bP,T}\pinv{\bS_T}.
\end{align}
Finally, the data symbols $\bS_D$ can be detected using, e.g., the well-known LMMSE detector:
\begin{align}
	\hat\bS_D = \inv{( \herm{\hat{\bH}_\bP}\hat{\bH}_\bP + \No\bI_U )} \herm{\hat{\bH}_\bP} \bar\bY_{\bP,D}. \label{eq:p_lmmse}
\end{align}

\subsection{LMMSE Equalization} \label{sec:mash-lmmse}
The orthogonal projection method from the previous subsection is highly effective. However, 
it requires the computation of a singular value decomposition (of $\bar{\bY}_J$) and the explicit 
estimation of the jammer interference dimension $\Ie$\!.
Both of these can be avoided if, instead of an orthogonal projection, we directly use
an LMMSE-type equalizer based on an estimate of the jammer's spatial covariance matrix
$\bC_{J}\triangleq\Ex{}{\bJ\bar{\bW}\herm{(\bJ\bar{\bW})}}$. 
Such an estimate can be obtained from $\bar{\bY}_J$ in \eqref{eq:j_train} as 
\begin{align} \textstyle
	\hat\bC_J = \frac{1}{R} \bar{\bY}_J\herm{\bar{\bY}_J}. 
\end{align}
An ``LMMSE-type'' jammer-mitigating estimate  
of $\bH$ based on \eqref{eq:pilots} is then
(the impact of the thermal noise $\bar\bN_T$ is neglected) 
\begin{align}
	\hat\bH = \Big(\bI_B + \frac{1}{U} \hat\bC_J\Big)^{-1} \bar{\bY}_T\pinv{\bS_T}, \label{eq:mit_chest}
\end{align}
and the LMMSE estimate of the data $\bS_D$ based on \eqref{eq:data} is
\begin{align}
	\hat\bS_D = \herm{\hat\bH}\big(\hat\bH\herm{\hat\bH}  + \No\bI_B + \hat\bC_J\big)^{-1} \bar{\bY}_D. \label{eq:mit_det}
\end{align}
This avoids the calculation of an SVD, but would require~us~to invert two matrices of size $B\times B$ 
(in \eqref{eq:mit_chest} and in \eqref{eq:mit_det}).
However, using identities from \cite{petersen12a}, we can rewrite \eqref{eq:mit_chest} and \eqref{eq:mit_det} 
as
\begin{align}
	\hat\bH = \big(\bI_B - \bar{\bY}_J\inv{\big(UR\,\bI_R + \herm{\bar{\bY}_J}\bar{\bY}_J\big)}\herm{\bar{\bY}_J} \big)
	\bar{\bY}_T\pinv{\bS_T} \label{eq:small_chest}
\end{align}
and 
\begin{align}
	\! \bA &= \!\bigg( \!\No\bI_{U+R} + \begin{bmatrix} \herm{\hat\bH}\\ \frac{1}{\sqrt{R}}\herm{\bar{\bY}_J}\end{bmatrix}\!
		\begin{bmatrix} \hat\bH, & \!\!\!\!\!\frac{1}{\sqrt{R}}\bar{\bY}_J \end{bmatrix} \!\bigg)^{\!\!-1}\! 
		\begin{bmatrix} \herm{\hat\bH}\\ \frac{1}{\sqrt{R}}\herm{\bar{\bY}_J}\end{bmatrix} \label{eq:small_det}\! \\
	\! \hat\bS_D \!&= \bA_{(1:U)}\bar{\bY}_D.
\end{align}
So we only need to invert an $R\times R$ matrix (in \eqref{eq:small_chest}) and a $(U+R)\times(U+R)$ matrix
(in \eqref{eq:small_det}).
In contrast, the orthogonal projection approach requires computing the SVD of a $B\times L$ matrix 
and the inverting a $U\times U$ matrix (in \eqref{eq:p_lmmse}).

\subsection{Joint Jammer Mitigation and Data Detection (JMD)} \label{sec:mash-maed}

JMD is a recent paradigm for smart jammer mitigation.
In JMD, the jammer subspace is estimated and nulled (through an orthogonal projection) \emph{jointly}
with detecting the transmit data over an entire transmission frame \cite{marti2023maed, marti2023jmd}. 
JMD operates by approximately solving a nonconvex optimization problem.
In principle, this can provide excellent performance \emph{even without} any jammer training period at all 
(corresponding to $R=0$). 
However, JMD suffers from several limitations:
(i) it performs poorly against jammers with very short duty cycle, 
(ii) it needs to know the dimension $\Ie$ of the jammer interference, and 
(iii)~it does not reliably converge to the correct solution if $\Ie$ is large.
All of these shortcomings can be substantially alleviated if JMD is combined with MASH:
(i)~MASH transforms jammers with short duty cycles into fully non-sparse barrage jammers, 
(ii)~in combination with non-zero redundancy ($R>0$), $\bar{\bY}_J$ can be used to estimate the dimension 
of the jammer interference (cf. Footnote~\ref{footnote:est_Ie}), and
(iii)~$\bar{\bY}_J$ can be used for optimal initialization to improve performance against high-dimensional jammers.

We exemplarily write down the optimization problem to solve if the JMD method MAED \cite{marti2023maed} 
is enhanced with MASH (a detailed account will be given in an extended journal paper): 
\begin{align} \textstyle
	\min_{\tilde\bP, \tilde\bH_{\bP}, \tilde\bS_D}
	\big\|\tilde\bP[\bY_{\bP,T},\bY_{\bP,D}] - \tilde\bH_{\bP}[\bS_T,\tilde\bS_D]\big\|_F^2.
\end{align}
Here, the range of $\tilde\bP$, $\tilde\bH_\bP$, and $\tilde\bS_D$ are the $(B-\Ie)$-dimen-sional Grassmanian manifold, 
$\opC^{B\times U}$, and $\setS^{U\times D}$, respectively. 
MASH allows $\Ie$ to be estimated from $\bar{\bY}_J$ (cf. Footnote~\ref{footnote:est_Ie}), 
and $\tilde\bP, \tilde\bS_D,$ and $\tilde\bH_\bP$ can be initialized using the 
orthogonal projection method of \fref{sec:mash-pos}, which results in high robustness
even against many-antenna jammers.

\section{Experimental Evaluation} \label{sec:eval}

\subsection{Simulation Setup} \label{sec:setup}

We simulate the uplink of a massive MU-MIMO system with $B=64$ antennas at the BS and $U=16$ UEs. 
The channel vectors are generated using QuaDRiGa \cite{jaeckel2014quadriga} with the 
3GPP 38.901 urban macrocellular (UMa) channel model~\cite{3gpp22a}. 
The carrier frequency is $2$\,GHz and the BS antennas are arranged as a uniform linear array (ULA)
with half-wavelength spacing. The UEs and the jammer are distributed randomly at distances 
from $10$\,m to $250$\,m in a $120^\circ$ angular sector in front of the BS, with a minimum angular 
separation of $1^\circ$ between any two UEs as well as between the jammer and any UE. 
The jammer can be a single- or a multi-antenna jammer (see \fref{sec:jammers}).
The antennas of multi-antenna jammers are arranged as a half-wavelength ULA that is 
facing the BS's direction. All antennas are omnidirectional. 
The UEs use $\pm3$\,dB power~control. 

As in \fref{sec:mash}, we consider coherent data transmission. 
The framelength is $L=100$, and is divided into a redundancy of $R=16$, 
$T=16$ pilot samples, and $L-R-T=68$ data samples. 
The transmit constellation $\setS$ is QPSK, and the pilots $\bS_T$ 
are chosen as a $16\times16$ Hadamard matrix (normalized~to unit symbol energy).
We characterize the strength of the jammer interference relative to the 
strength of the average UE via
\begin{align} \textstyle
	\rho \define \frac{\|\bJ\bW\|_F^2}{\frac1U\Ex{\bS}{\|\bH\bX\|_F^2}}.
\end{align}
The average signal-to-noise ratio (SNR) is defined as
\begin{align}
\textit{SNR} \define \frac{\Ex{\bS}{\|\bH\bX\|_F^2}}{\Ex{\bN}{\|\bN\|_F^2}}.
\end{align}
As performance metrics, we use uncoded bit error rate (BER) as well as 
a surrogate for error vector magnitude (EVM) \cite{3gpp21a} which we call modulation~error ratio (MER) 
and define as
\begin{align} 
	\textit{MER}\triangleq \mathbb{E}\big[\|\hat\bS_D - \bS_D\|_F\big]\big/\mathbb{E}\big[\bS_D\|_F\big].
\end{align}

\subsection{Methods and Baselines}
\begin{figure}[tp]
\centering
\subfigure[MASH]{\includegraphics[height=1.05cm]{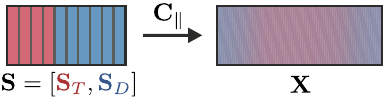}\label{fig:mash}}
\hfill
\subfigure[Baselines]{\includegraphics[height=1.05cm]{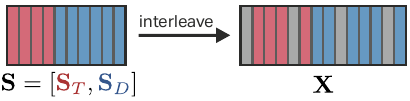}\label{fig:baselines}}
\caption{MASH (left) maps a length-$K$ signal $\bS=[\bS_T,\bS_D]$ to a length-$L$ signal $\bX$ by multiplying
with $\Cpar$. The baselines (right) map a length-$K$ signal $\bS=[\bS_T,\bS_D]$ to a length-$L$ signal 
by interleaving it with evenly distributed zero-symbols (depicted in gray) that serve as jammer training period.}
\label{fig:mash_vs_baselines}
\vspace{-2mm}
\end{figure}

We compare MASH against baselines which, instead of embedding the pilots
and data symbols in a secret higher-dimensional space, transmit them in the conventional
way, but interleave them with $R$ zero-symbols that are evenly distributed over the frame
and serve as jammer training period (cf. \fref{fig:mash_vs_baselines}).

\begin{figure*}[tp]
\centering
\!\!\!\!\!\!
\subfigure[single-antenna barrage jammer \tinygraycircled{1}]{
\includegraphics[height=3.25cm]{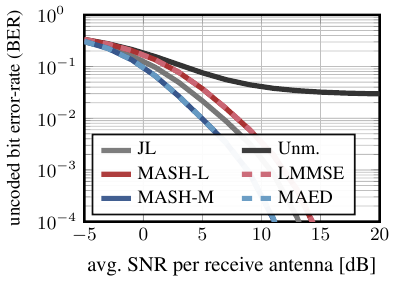}
\label{fig:barrage}
}\!\!
\subfigure[single-antenna data jammer \tinygraycircled{2}]{
\includegraphics[height=3.25cm]{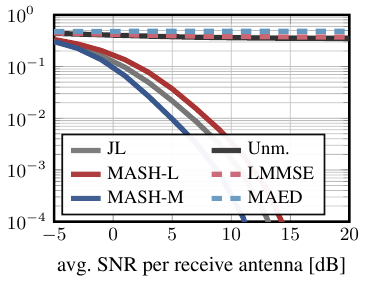}
\label{fig:data}
}\!\!
\subfigure[single-antenna pilot jammer \tinygraycircled{3}]{
\includegraphics[height=3.25cm]{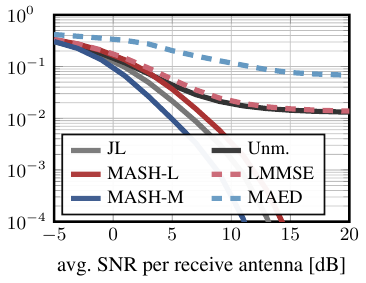}
\label{fig:pilot}
}\!\!
\subfigure[single-antenna sparse jammer \tinygraycircled{4}]{
\includegraphics[height=3.25cm]{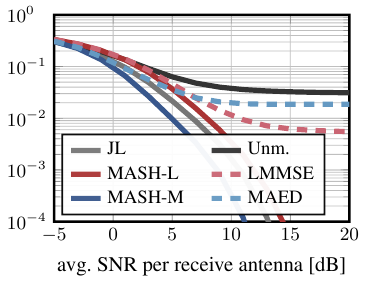}
\label{fig:sparse}
}\!\!\!\!
\!\!\!\!\!\!
\subfigure[multi-ant. eigenbeamf. jammer \tinygraycircled{5}]{
\includegraphics[height=3.25cm]{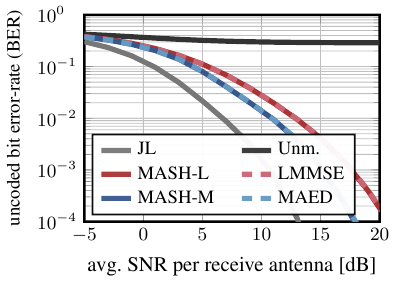}
\label{fig:eigenbeamforming}
}\!\!
\subfigure[multi-antenna data jammer \tinygraycircled{6}]{
\includegraphics[height=3.25cm]{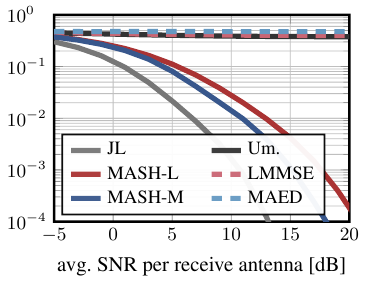}
\label{fig:multi-data}
}\!\!
\subfigure[multi-ant. dynamic jammer \tinygraycircled{7}]{
\includegraphics[height=3.25cm]{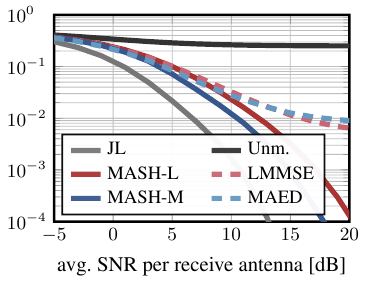}
\label{fig:dynamic}
}\!\!
\subfigure[multi-antenna repeat jammer \tinygraycircled{8}]{
\includegraphics[height=3.25cm]{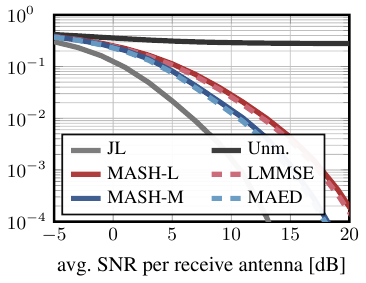}
\label{fig:repeater}
}\!\!\!\!
\caption{
Bit error rate (BER) performance of the different MASH implementations and baselines for eight different types of jammers. 
}
\label{fig:jammers}
\vspace{-2mm}
\end{figure*}

\subsubsection*{\textbf{MASH-L}}
This receiver operates as described in Sec. \ref{sec:mash-lmmse}.

\subsubsection*{\textbf{MASH-M}}
This receiver operates as described in Sec. \ref{sec:mash-maed}. 

\subsubsection*{\textbf{JL}}
This baseline operates on a jammerless (JL) system. The transmitter operates as
depicted in \fref{fig:baselines}. The receiver ignores
the receive samples of the jammer training period and performs LS 
channel estimation and LMMSE data detection. 

\subsubsection*{\textbf{Unmitigated (Unm.)}}
This baseline does not mitigate the jammer. The transmitter operates as
depicted in \fref{fig:baselines}. The receiver ignores
the receive samples of the jammer training period and performs LS 
channel estimation and LMMSE data detection as one would in a jammerless environment.

\subsubsection*{\textbf{LMMSE}}
This baseline is identical to MASH-L, except that it does not use secret subspace embeddings. 
The transmitter operates as depicted in \fref{fig:baselines}.
The receiver estimates the jammer's spatial covariance matrix
using the receive samples $\bY_J$ from the training period, 
$\hat\bC_J = \frac1R \bY_J\herm{\bY_J}$, 
and then performs jammer-mitigating channel estimation and data detection analogous to 
\eqref{eq:mit_chest} and \eqref{eq:mit_det}. 

\subsubsection*{\textbf{MAED}}
This baseline is identical to MASH-M, except that it does not use secret subspace embeddings. 
The transmitter operates as depicted in \fref{fig:baselines}. The receiver uses JMD and operates as 
described in \cite{marti2023maed}.

For the sake of figure readability, we omit the orthogonal projection variant of \fref{sec:mash-pos}
from our experiments, which performs similar to MASH-L. 
None of the methods is given a~priori knowledge about the dimension $\Ie$ of the interference space. 
The methods that require such knowledge estimate $\Ie$ as the number of the singular values of 
$\bar\bY_J$~(\mbox{MASH-M}) or of~$\bY_J$ (MAED) that exceed $2\sqrt{B\No}$ (cf. Footnote~\ref{footnote:est_Ie}).

\subsection{Jammers} \label{sec:jammers}
The transmit signals of all jammers are normalized to \mbox{$\rho=30\,$dB.}
All multi-antenna jammers have $I=10$ antennas. 
\subsubsection*{\tinygraycircled{1} Single-antenna barrage jammer} 
This jammer transmits i.i.d. $\setC\setN(0,1)$ symbols in all samples.
\subsubsection*{\tinygraycircled{2} Single-antenna data jammer}
This jammer transmits i.i.d. $\setC\setN(0,1)$ symbols 
in all samples in which the UEs transmit data symbols\footnote{This refers to the samples of 
the baseline methods as in \fref{fig:baselines}. The subspace embeddings of MASH entail that 
data symbols are not transmitted during specific samples, cf. \fref{fig:mash}.
The same applies to jammers \tinygraycircled{3} and~\tinygraycircled{6}.}
and is idle in all other samples. 
\subsubsection*{\tinygraycircled{3} Single-antenna pilot jammer}
This jammer transmits i.i.d. $\setC\setN(0,1)$ symbols 
during the UE pilot transmission period. It is idle in all other samples.
\subsubsection*{\tinygraycircled{4} Single-antenna sparse jammer}
This jammer transmits i.i.d. $\setC\setN(0,1)$ symbols 
in a fraction of $\alpha=0.1$ of \mbox{non-contigu}-ous randomly selected samples and is idle in all other~samples.
\subsubsection*{\tinygraycircled{5} Multi-antenna eigenbeamforming jammer}
This jammer is assumed to have full knowledge of $\bJ=\bsfU\bsfSigma\herm{\bsfV}$ and 
uses eigenbeamforming to transmit $\bW=\bsfV\tilde\bW$, where the entries of $\tilde\bW$
are i.i.d. $\setC\setN(0,1)$. By Definition~\ref{def:barrage}, this jammer is
considered a barrage jammer.
\subsubsection*{\tinygraycircled{6} Multi-antenna data jammer}
This jammer transmits i.i.d. random vectors $\bmw_k\sim\setC\setN(\mathbf{0},\bI_I)$ for all samples $k$ in which 
the UEs transmit data symbols and is idle in all other~samples. 
\subsubsection*{\tinygraycircled{7} Multi-antenna dynamic-beamforming jammer}
At any given instance $k$, this jammer uses only a subset of its antennas, but it 
uses dynamic beamforming to change its antennas over time. 
Specifically, its time-$k$ beamforming matrix $\bA_k$ contains at most eight non-zero rows 
(the index set of these rows is generated uniform at random) whose entries are drawn i.i.d.~at random from $\setC\setN(0,1)$.
The matrix $\bA_{k+1}$ is equal to $\bA_k$ with probability $0.95$, and with probability
$0.05$, it is redrawn at random. The vectors $\tilde\bmw_k$ are drawn from 
$\setC\setN(\mathbf{0},\bI_I)$ for all~$k$. 
\subsubsection*{\tinygraycircled{8} Multi-antenna repeat jammer}
This jammer is assumed to have sensing capabilities that allow it to perfectly detect the UE transmit signal. 
The jammer then simply repeats the transmit signal $\bX_{(1:i)}$ (cf. \eqref{eq:embed}) of the first $I$ UEs with a delay of 
one sample, i.e., $\bW = [\mathbf{0}_{I\times 1},\bX_{(1:I),[1:L-1]}]$.

\subsection{Results}
\begin{figure}
\vspace{-2mm}
\centering
\!\!\!\!\!\!\!\!
\subfigure[jammer \tinygraycircled{1}]{
\includegraphics[height=3.25cm]{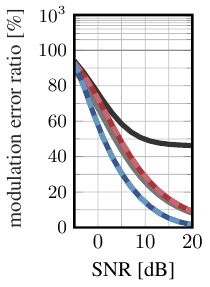}
}\!\!\!\!
\subfigure[jammer \tinygraycircled{2}]{
\includegraphics[height=3.25cm]{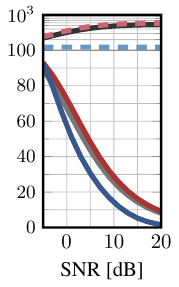}
}\!\!\!\!
\subfigure[jammer \tinygraycircled{3}]{
\includegraphics[height=3.25cm]{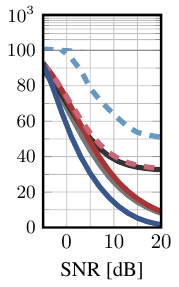}
}\!\!\!\!
\subfigure[jammer \tinygraycircled{4}]{
\includegraphics[height=3.25cm]{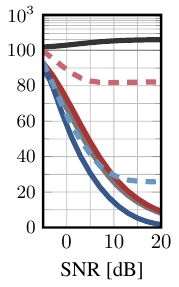}
}\!\!\!\!

\!\!\!\!\!\!\!\!
\subfigure[jammer \tinygraycircled{5}]{
\includegraphics[height=3.25cm]{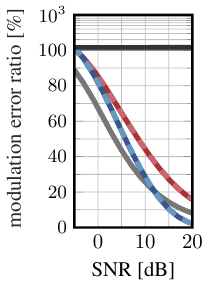}
}\!\!\!\!
\subfigure[jammer \tinygraycircled{6}]{
\includegraphics[height=3.25cm]{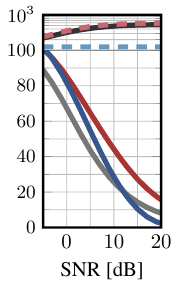}
}\!\!\!\!
\subfigure[jammer \tinygraycircled{7}]{
\includegraphics[height=3.25cm]{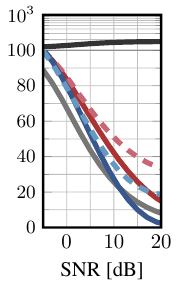}
}\!\!\!\!
\subfigure[jammer \tinygraycircled{8}]{
\includegraphics[height=3.25cm]{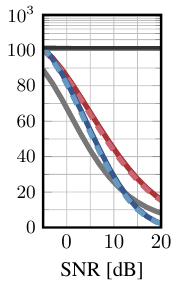}
}\!\!\!\! 
\caption{
Modulation error ratio (MER) performance of the MASH implementations and baselines for the eight 
types of jammers. The correspondence between line styles and methods is the same as in \fref{fig:jammers}.
Note that the MER can exceed $100\%$ when mitigation fails. Note also that 
the $y$-axis is scaled linearly between $0$ and $100$, and logarithmically between $100$ and $1000$.
\label{fig:mers}
\vspace{-3mm}
}
\end{figure}

The results in \fref{fig:jammers}, \fref{fig:mers} show that MASH mitigates \emph{all} jammers under consideration, 
and so confirm the universality of MASH. We start by discussing the BER results of \fref{fig:jammers}.

When facing the single-antenna barrage jammer in \fref{fig:barrage}, 
MASH-L and MASH-M have exactly the same performance as their non-MASH counterparts LMMSE and MAED. 
This is expected, since transforming a barrage jammer into its equivalent barrage jammer changes nothing.
LMMSE and~MASH-L perform close to the JL baseline.
MAED and \mbox{MASH-M} use a more complex, nonlinear JED-based data detector~\cite{marti2023maed}, and so are able to 
perform even better than the JL baseline.

For the more sophisticated single-antenna jammers, however, the picture changes (\fref{fig:pilot}--\fref{fig:sparse}): 
LMMSE performs poorly against all of them, since the training receive matrix~$\bY_J$ does not (or not~necessarily, 
in the case of jammer \tinygraycircled{4}) contain samples in which the jammer is transmitting. 
MAED performs just as bad.\footnote{In principle, MAED would be able to mitigate such jammers---\emph{if} 
given knowledge of $\Ie$ \cite{marti2023maed}. Here, the problem is that MAED's estimate of $\Ie$ will (wrongly) be zero whenever $\bY_J$ 
contains no jammer samples.}
However, MASH-L and \mbox{MASH-M} transform all these jammers into their equivalent barrage~jammers and thus---as
predicted by theory---have \emph{exactly} the same performance as they do for the barrage jammer in \fref{fig:barrage}.

For the multi-antenna jammers, we observe the following (\fref{fig:eigenbeamforming}--\fref{fig:repeater}):
The eigenbeamforming jammer in \fref{fig:eigenbeamforming} is also a barrage jammer (even if it uses beamforming). 
So, as in \fref{fig:barrage}, the MASH methods have exactly the same performance
as their non-MASH counterparts. However, since the jammer now occupies 10 spatial dimensions (instead of~1) 
that need to be suppressed, the performance gap between the mitigation methods and the JL baseline (which does
not have to suppress any dimensions) is larger than in \fref{fig:barrage}
(the JMD methods MASH-M and MAED now perform worse than JL in spite of their nonlinear data detectors).
For the multi-antenna data jammers in \fref{fig:multi-data}, the non-MASH baselines fail spectacularly, 
while the MASH methods still have the same performance as for the multi-antenna barrage jammer in \fref{fig:eigenbeamforming}.
Similar obervations apply to the dynamic-beamforming jammer in \fref{fig:dynamic}, except that here, 
the non-MASH baselines can estimate parts of the interference subspace (or all of it, if they are lucky) 
during the training period and thus fare somewhat better. 
Finally, the repeat jammer in \fref{fig:repeater} behaves functionally almost like a barrage jammer and is sucessfully 
mitigated by all mitigation methods 
(note that the jammer transmits different signals for the MASH and the non-MASH methods). 
This experiment shows that sneaky repeat attacks are not able 
to overcome MASH. This was to be expected, of course, since $\text{row}(\Cpar)$ is, in general, not closed 
under cyclic shifts of the rows of $\Cpar$. 
(However, to prevent repeat attacks against MASH on the level of entire frames, the transmitters and receiver should 
update the matrix~$\bC$ according to a pseudo-random sequence after every transmission frame.)

The MER results in \fref{fig:mers} show strong agreement with the BER results in \fref{fig:jammers}.
The \emph{only} observable difference is that the nonlinear methods MAED/MASH-M perform slightly better compared the 
linear methods
JL/LMMSE/MASH-L in terms of MER than in terms of BER. (This is due to the data symbol 
prior of MAED/\mbox{MASH-M}, 
which pulls symbol estimates to constellation points and so achieves lower euclidean error than linear detection
even when the underlying decoded bits are the same.) 
Even so, the results of \fref{fig:mers} agree with those of \fref{fig:jammers}: the MASH methods are the only 
ones that successfully mitigate \emph{all} types of jammers, while the traditional 
baselines that do not use subspace embeddings fail in many of the cases.

\section{Conclusions}
We have provided a mathematical definition to capture the essence of the notion of barrage jammers, 
which are easy to mitigate using MIMO processing. We  have then proposed MASH, a novel method where
the transmitters embed their signals in a secret temporal subspace out of which the receiver raises them, 
thereby provably transforming \emph{any} jammer into a barrage jammer. Considering a massive MU-MIMO uplink example scenario, 
we have provided three concrete variations of MASH with different performance-complexity tradeoffs. 
Numerical simulations have confirmed that these MASH-based methods are able to mitigate all jammers under consideration.

\appendix[Proof of \fref{thm:barrage}] \label{app:app}
We rewrite the compact SVD of $\bJ\bW = \bsfU\bsfSigma\herm{\bsfV}$ as a full SVD
$\bJ\bW = \bU\bSigma\herm{\bV}$ with $\bU\in\opC^{B\times B}, \bSigma\in\opC^{B\times L},$ and $\bV\in\opC^{L\times L}$. 
We then have $\bJ\bW\herm{\bC} = \bU\bSigma\herm{(\bC\bV)}$.
Since~$\bC$ is Haar distributed, so is $\bC\bV$ \cite[Thm.\,1.4]{meckes2019random}. 
The compact SVD of $\bJ\bW\herm{\bC}\!=\!\bar{\bsfU}\bar{\bsfSigma}\herm{\bar{\bsfV}}$ is 
$\bar{\bsfU} \!=\! \bU_{[1:\Ie]}=\bsfU$, \mbox{$\bar{\bsfSigma}\!=\!\bSigma_{(1:\Ie),[1:\Ie]}=\bsfSigma$,}
which implies $\bar{\bsigma}=\bsigma$, and 
$\bar{\bsfV} = (\bC\bV)_{[1:\Ie]}$. Since $\bC\bV$ is Haar distributed, the columns of $\bar{\bsfV}$ are
distributed uniformly over the complex $L$-dimensional sphere. $\hfill\largesecret$

\balance


\end{document}